# IR radiometer sensitivity and accuracy improvement by eliminating spurious radiation for emissivity measurements on highly specular samples in the 2–25 μm spectral range

T. Echániz [a], R.B. Pérez-Sáez [a,b], M.J. Tello [a,b]

[a] Departamento de Física de la Materia Condensada, Facultad de Ciencia y Tecnología, Universidad del País Vasco, Barrio Sarriena s/n, 48940 Leioa, Bizkaia, Spain

[b] Instituto de Síntesis y Estudio de Materiales, Facultad de Ciencia y Tecnología, Universidad del País Vasco, Apdo. 644, 48080 Bilbao, Bizkaia, Spain

Abstract

This paper indicates that the sensitivity and accuracy of an infrared radiometer for emissivity measurements depends not only on its design and the measurement method, but also on the spurious radiation. This spurious radiation must be taken into account in the calibration processes since it can be of the same order of magnitude as that of the sample in highly reflective surfaces. Its presence may also be the cause of the inability to detect small surface emissivity changes induced by any surface or bulk properties (anomalous skin effect, phase transitions, etc.). In this paper, the analysis of the spurious radiation is performed for a T-form radiometer and a measurement method where the surroundings are considered to emit as if they were black bodies. However, the results and conclusions that are obtained can be extended without difficulty to any type of radiometer and to all measurement methods. Our research shows that if the sample is placed normal to the emitted radiation optical path, two different spurious radiation sources are detected. However, for low emitting and highly specular materials, they can be eliminated by tilting the sample between 6 and 20° without modifying the spectral emissivity.

1. Introduction

In a large number of industrial applications (solar energy, nuclear fusion reactors, chemical reactors, machining tools, aerospace industry, etc.) as well as in basic science, it is necessary to have accurate surface emissivity values of the materials and to know the dependence on temperature, wavelength, emission angle and atmosphere $\varepsilon(\lambda,T)$ [1–6]. The emissivity also strongly depends on the surface state (roughness, cleanliness, adsorption, absorption, impurities, mechanical stresses, etc.) [7–9]. In addition, the electromagnetic theory defines a direct relation between the emission spectrum of a crystalline homogeneous smooth material with its optical and dielectric properties [10,11].

The spectral emissivity $\varepsilon(\lambda,T)$ is obtained using direct methods that require designing a radiometer. The emissivity is usually obtained by comparing the sample and blackbody spectra at the same temperature and under similar geometric and optical conditions [10,11]. Several radiometer designs and different measurement methods (usually determined by the radiometer design) have been reported. See for example [12–25] and references therein. In this paper the theoretical equations refer to the blacksur method [26]. A T-form radiometer is used in our laboratory and its geometrical configuration is shown in Fig. 1 [13]. Its four main parts are the sample chamber, the blackbody, the optical entrance and the FT-IR spectrometer with the detector inside. The detector is a DTLaGS detector that works in the 1.43–25 µm spectral range. The chamber possesses a double metallic layer in order to insert water and refrigerate the chamber and the sample surroundings to room temperature. The chamber is painted with a black paint (Nextel Velvet Coating 811-21) so it can emit as a blackbody and therefore possible reflections inside the chamber are eliminated [13]. The sample temperature is measured by means of two K-type thermocouples on the selected sample area where good temperature homogeneity is ensured. The radiometer allows measuring between 100 and 1000 °C. In any case, the conclusions of this paper are valid for all measurement methods and all radiometer types.

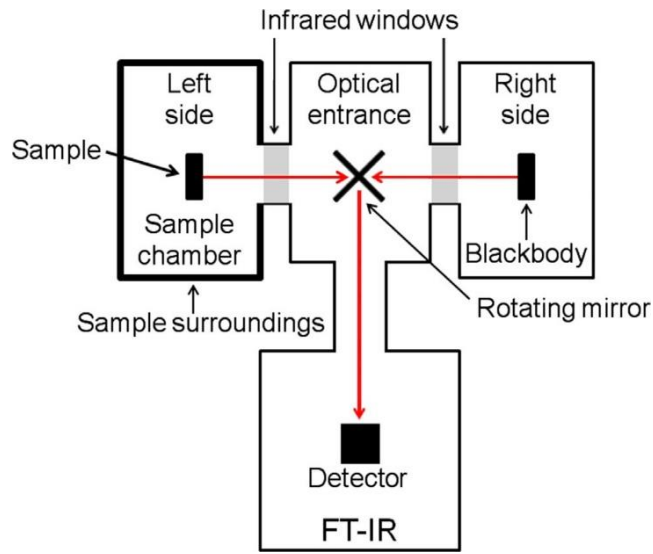

Fig. 1. Schematic view of a T-form radiometer.

2. Theoretical background

The accuracy, the calibration and the uncertainty of the most common measurement methods were analyzed in the past. The experimental uncertainty is 2% between 4 and 12 µm, and it reaches its maximum value of 6% at 25 µm. See [26–33] and references therein. In this paper the blacksur method is used because the literature suggests that this method, in which the surrounding radiation is considered to be equivalent to a blackbody radiation, gives very low errors. In addition, when a large sample chamber is used, these errors are practically independent of the surroundings.

In the blacksur method the radiation per unit wavelength and solid angle leaving the sample surface $L_s^*(\lambda, T_s)$ at temperature $T_s$ is related to the sample spectral emissivity $\varepsilon_s(\lambda, T_s)$ by the following equation [26]:

$$L_s^*(\lambda, T_s) = \varepsilon_s(\lambda, T_s) L(\lambda, T_s) + [1 - \varepsilon_s(\lambda, T_s)] L(\lambda, T_{sur}), \qquad (1)$$

where $L(\lambda,T_s)$ and $L(\lambda,T_{sur})$ are the spectral directional emission intensities given by Planck's equation at the sample (s) and surrounding (sur) temperatures respectively. It should be noted that in Eq. (1) the radiation emitted by the surroundings and then reflected in the sample is taken into account. However, the sample emissivity cannot be obtained from Eq. (1) because the radiation reaching the radiometer detector $S_s(\lambda,T_s)$ is not equal to the radiation leaving the surface of the sample $L_s^*(\lambda,T_s)$. Both are connected by the following equation:

$$S_s(\lambda, T_s) = R(\lambda)\left[A_s F_{s-det} L_s^*(\lambda, T_s) + L_0(\lambda)\right], \qquad (2)$$

where $A_s$ is the sample area, $F_{s\text{-}det}$ is a configuration factor [10,11], $R(\lambda)$ is the response function of the radiometer and $L_0(\lambda)$ refers to the background radiation that reaches the detector directly [26–28]. $L_0(\lambda)$ is usually considered to be mainly composed by radiation emitted by the inner parts of the radiometer. In order to eliminate the temperature dependence of the last two parameters ($R(\lambda)$ and $L_0(\lambda)$) the radiometer temperature must be controlled. Usually, the temperature control is performed with a fluid layer around the lateral walls of the sample chamber together with a laboratory temperature control. The values of the response function, the background radiation and the shape factor are obtained in the radiometer calibration. For this purpose it is necessary to use a blackbody. It should also be noted that the IR window transmittance $T_{IR}$ is implicitly included in the response function. By analogy with Eq. (2), the blackbody radiation reaching the radiometer detector can be written as:

$$S_{bb}(\lambda, T_{bb}) = R(\lambda)[A_{bb} F_{bb-det} L(\lambda, T_{bb}) + L_0(\lambda)], \qquad (3)$$

where $A_{bb}$ is the blackbody area, $F_{bb\text{-}det}$ is the blackbody-detector configuration factor and $L_0(\lambda)$ is the spectral directional emission intensity given by Planck's equation at $T_{bb}$. The radiometer design should ensure that the surface areas of the two emitting sources, the blackbody and the sample ones, viewed by the detector, are equal and that the radiation emitted from both surfaces has the same optical path length to reach the detector. Then, $R(\lambda)A_s F_{s\text{-}det} = R(\lambda)A_{bb} F_{bb\text{-}det} = R^*(\lambda)$ is satisfied. $R^*(\lambda)$ and $L_0(\lambda)$ can be obtained from Eq. (3) by measuring the radiation of a blackbody for two temperatures. Then, the equation to obtain the spectral emissivity of the sample from the experimental measurements is obtained using the Eqs. (1)–(3); being the same as the one in reference [26]:

$$\varepsilon_s(\lambda, T_s) = \frac{\frac{[S_s(\lambda,T_s)-S_{bb1}(\lambda,T_{bb1})][L(\lambda,T_{bb1})-L(\lambda,T_{bb2})]}{[S_{bb1}(\lambda,T_{bb1})-S_{bb2}(\lambda,T_{bb2})]} + L(\lambda, T_{bb1}) - L(\lambda, T_{sur})}{L(\lambda, T_s) - L(\lambda, T_{sur})}, \qquad (4)$$

where the bb1 and bb2 subindexes refer to the two blackbody temperatures.

Eq. (4) is usually used to obtain the emissivity spectra for each temperature. However, a more careful analysis of spurious radiation sources indicates that to increase the accuracy and sensitivity of the radiometer two other radiation sources that also reach the radiometer detector and must be removed to get the sample emissivity value should be taken into account. In some cases, the emissivity value of this spurious radiation is of the same order of magnitude of the sample emissivity. This occurs when the surfaces are highly reflective like many metals. These spurious radiations also affect substantially the emissivity measurement when we want to detect properties or effects that produce small changes in the sample emissivity. A good example is the detection of the anomalous skin effect [34,35].

The first of these spurious radiations is emitted by the radiometer detector, which reaches the surface of the sample by inverting the sample-detector optical path. This radiation that is reflected by the sample returns to the detector and thus increases the value of the radiation reaching the detector. On the other hand, a small part of the radiation emitted by the sample is reflected by the infrared window. Afterwards, it is also reflected on the sample and finally reaches the detector (Fig. 1). One could consider multiple reflections (two or more), but their values are negligible and below the experimental uncertainty. These spurious signals are partially damped due to the increment of the optical path and their interaction with lenses and mirrors. To take account of these new sources of radiation, Eq. (1) must include two terms that account for these new spurious signals. Besides, according to Fig. 2a, when the emitting surface is nearly specular and perpendicular to the optical path, the emission coming from the surroundings can be neglected because no surrounding radiation reflected on the sample is able to reach the detection system.

With these assumptions Eq. (1) becomes:

$$L_s^*(\lambda, T_s) = \varepsilon_s(\lambda, T_s)L(\lambda, T_s) + \varepsilon_{det}(\lambda, T_{det})\mathcal{T}_{IR}F_{opt}[1 - \varepsilon_s(\lambda, T_s)]L(\lambda, T_{det}) \\ + \varepsilon_s(\lambda, T_s)L(\lambda, T_s)\mathcal{R}_{IR}[1 - \varepsilon_s(\lambda, T_s)], \quad (5)$$

where $T_{IR}$ and $R_{IR}$ are the infrared window transmittance and reflectance respectively (the absorptance is neglected) [36], the det subindex refers to the radiometer detector and $F_{opt}$ is a signal loss factor caused by the optical components. Both $F_{opt}$ and $T_{IR}$ are also included in the response function $R^*(\lambda)$ but are considered part of Eq. (5) since the radiation of the detector passes through this components twice instead of once before it is detected.

Eq. (5) is difficult to use due to the lack of knowledge of some physical parameters associated to the spurious signal, such as the detector emissivity and its temperature. However, for specular or nearly specular samples, Lambert's cosine law allows, with a small tilt of the sample, eliminating spurious radiation leaving invariant the value of the emissivity of the sample [10]. Here, the changes occurring in detected radiation sources when the sample is tilted must be considered. Fig. 2b shows that the two spurious contributions are absorbed in the sample surroundings after they have been reflected by the sample and therefore can be eliminated from Eq. (5). Besides, with this configuration, radiation incoming from the sample chamber will be detected after being reflected by the sample and Eq. (4) can be used again to calculate the emissivity. Obviously, rotation can be applied if the emission spectrum does not change with the angle. According to Lambert's cosine law this occurs in metals for emission angles below a = 20° [10]. This behavior can be experimentally verified or calculated from the refractive indexes. For example, using the copper literature refractive index shows that the difference in emissivity between a = 0° and a = 10° is below 0.01% in the 1–25 μm wavelength range at room temperature.

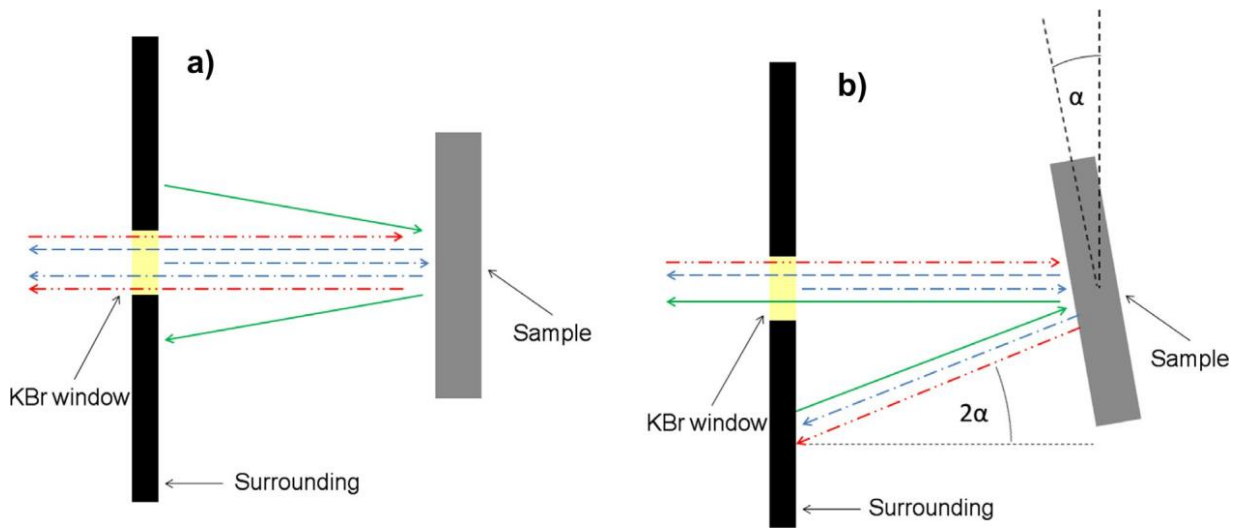

Fig. 2. Schematic representation of the radiation emitted by the sample surface and the spurious emissions for two cases: (a) radiation with the sample at normal position and (b) radiation with the sample tilted at an angle a. The long-dashed arrows (blue) and dotted-dashed (blue) ones refer to the sample emitted radiation; the long-dashed arrows being transmitted light and the dotted-dashed ones the reflected radiation on the KBr window. The double dotted-dashed (red) arrows refer to the detector radiation whereas the solid (green) ones indicate the surrounding radiation. (For interpretation of the references to colour in this figure legend, the reader is referred to the web version of this article.)

3. Experimental results and discussion

To experimentally verify the results proposed previously, the spectral directional emissivity of various materials have been measured with the radiometer [13]. In this paper only the experimental results of a pure metal (Mg) and an alloy (Mg-47%Zn-4%Al) are shown as an example. Both samples show very low emissivity values and have and average roughness (Ra) of 0.25 and 0.11 lm respectively. They were well characterized in our laboratory within a research program on thermal storage materials [37]. The emissivity spectra of both samples show the classic behavior predicted by the electromagnetic theory and the results of this work are valid for any low emissivity metallic material.

Fig. 3 shows the angular dependence of the magnesium emissivity for $\lambda$ = 11.5 µm at 300 °C. Similar curves were obtained for the entire wavelength range. The presence of spurious contributions in the vicinity of the normal direction is evident. Furthermore, in this case these contributions are quite important when compared with the value of the magnesium emissivity (around 60% of the real value). The angular range (peak width) where spurious contributions appear is related with the size of the infrared window and possibly with other components that encompass the radiation optical path. Fig. 3 also shows how, in accordance with Lambert's law, the emissivity stays constant between a = 6° and a = 15°. This is an experimental confirmation of the fact that the normal emissivity can be measured with high precision within this angular range.

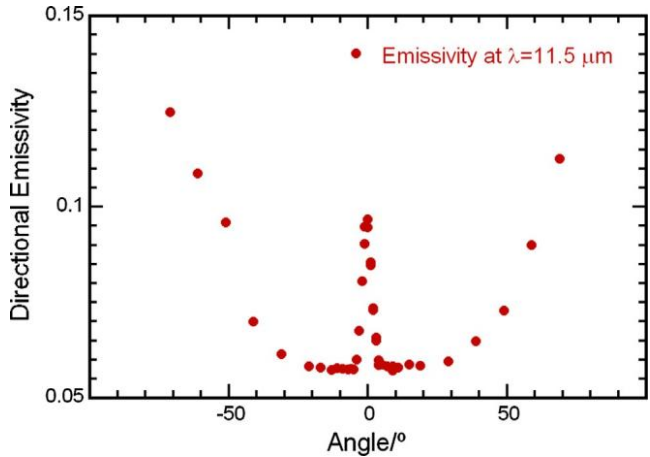

Fig. 3. Directional spectral emissivity of Mg at λ = 11.5 μm and T = 300 °C.

The spectral influence of the spurious radiation is shown in Fig. 4. In this figure the emission spectra obtained with Eq. (4) of the Mg-47%Zn-4%Al alloy for a = 0° and a = 10° and measured at 325 °C are plotted. It clearly shows that the spurious contribution to the value of the sample emissivity is very large (100% more at 17 lm) and becomes zero for λ <3 μm. Two factors seem to explain this dependence with the wavelength of the spurious signal. On one hand the detector radiation emission is almost non-existing at those wavelengths since the radiation incoming from the detector (the main spurious contribution), in this case a DLaTGS detector, is emitted at a much lower temperature (around 50 °C). On the other hand, the average height of the roughness pro- file (~1 lm) of this sample is close to the short wavelengths of the studied emissivity spectrum and therefore the surface is not as smooth as it is for longer wavelengths. As a consequence, the radiation at shorter wavelengths is not as specularly reflected as it is at longer ones and therefore less spurious signal should be detected.

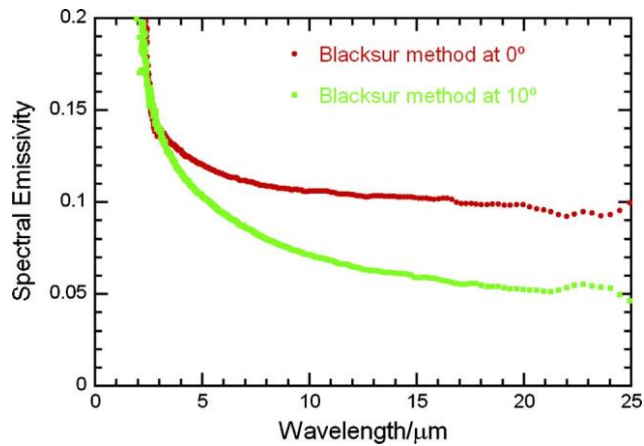

Fig. 4. Mg-47%Zn-4%Al emissivity at 0 and 10° for T = 325 °C.

Although we have eliminated the need to use the Eq. (5), it's interesting to complete the discussion with a more complete analysis of the detector emissivity. To do this, the Eqs. (1)–(3) allow obtaining the equation for the detector surface emissivity as a function of temperature:

$$\varepsilon_{det}(\lambda,T_{det})F_{opt}L(\lambda,T_{det}) = \frac{\frac{S_{0°}-S_{10°}}{[1-\varepsilon_s(\lambda,T_s)]R^*(\lambda)} - \varepsilon_s(\lambda,T_s)L(\lambda,T_s)\mathcal{R}_{KBr} + L(\lambda,T_{sur})}{\mathcal{T}_{KBr}}.$$

(6)

This equation allows weighing the detector spurious contribution. Fig. 5 does it by using the results from Fig. 4 and by giving values to the KBr infrared window reflectance and transmittance (0.083 and 0.9 respectively) [36,38]. Fig. 5 also displays the Planck curves for 52 and 60 °C to compare them to the detector radiation. In full accordance with what was indicated before, the emission intensity vanishes for short wavelengths. With this increase in the radiometer sensitivity structural phase transitions [39], as well as the anomalous skin effect in Cu have been observed [34,35].

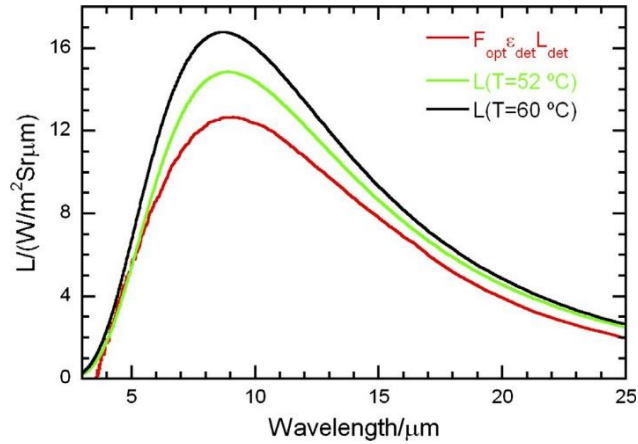

Fig. 5. Detector spectral emission intensity and Planck curves for T = 52 °C and T= 60 °C.

4. Conclusions

The influence of small spurious signals on the thermal emission spectrum has been analyzed in detail. It is found that the elimination of emissions of the detector surface and the reflections in the windows of the signal emitted by the sample significantly increases the radiometer accuracy. With a proper calibration and by eliminating the spurious signals the thermal emission spectrum becomes a high accuracy surface spectroscopic technique that allows measuring highly reflecting materials as well as very small variations in the material emissivity associated to different types of surface modifications. The tilting of the sample between 6 and 20° solves the problem of spurious radiations and permits using Eq. (4) [26] in order to obtain accurate emissivity values for low emitting samples.

Acknowledgements

T. Echániz would like to acknowledge the Basque Government its support through a Ph.D. fellowship.